\documentclass[twocolumn]{aastex61}
\usepackage{color}
\usepackage{multirow}
\usepackage{soul}
\usepackage{mathtools}

\newcommand{\U}[1]{\ensuremath{\mathrm{\ #1}}}
\newcommand{\UU}[2]{\ensuremath{\mathrm{\ #1^{#2}}}}

\received{August 21, 2018}
\revised{October 1, 2018}
\accepted{October 9, 2018}
\submitjournal{ApJL}

\shorttitle{TeV Gamma-Rays from the PSR~J2032+4127 Binary.}
\shortauthors{VERITAS and MAGIC Collaborations}

\newcommand{\psr}{PSR\,J2032+4127}
\newcommand{\be}{MT91\,213}
\newcommand{\psrbe}{PSR\,J2032+4127 / MT91\,213}
\newcommand{\tevj}{TeV\,J2032+4130}
\newcommand{\vername}{VER\,J2032+414} 
\newcommand{\magname}{MAGIC\,J2032+4127}

\newdimen \maxlen 
\newdimen \lenspace 
\newdimen \templen 
\newcommand{\setlen}[1]{\settowidth{\templen}{#1}
		\setlength{\lenspace}{\dimexpr(\maxlen - \templen)}}

\newcommand{\writeChitworow}[1]{\setlen{#1}$\begin{rcases*} \\ \\ 			\end{rcases*}$\hspace{\lenspace}#1}
\newcommand{\writeChithreerow}[1]{\setlen{#1}$\begin{rcases*} \\ \\ 		\\ \end{rcases*}$\hspace{\lenspace}#1}

\newcolumntype{Q}{>{\centering\arraybackslash}m{4cm}}

\begin{document}

\title{Periastron observations of TeV gamma-ray emission from a binary system with a 50-year period}
\author{A.~U.~Abeysekara} 
\affiliation{Department of Physics and Astronomy, University of Utah, Salt Lake City, UT 84112, USA}
\author{W.~Benbow} 
\affiliation{Fred Lawrence Whipple Observatory, Harvard-Smithsonian Center for Astrophysics, Amado, AZ 85645, USA}
\author{R.~Bird} 
\affiliation{Department of Physics and Astronomy, University of California, Los Angeles, CA 90095, USA}
\author{A.~Brill} 
\affiliation{Physics Department, Columbia University, New York, NY 10027, USA}
\author{R.~Brose} 
\affiliation{Institute of Physics and Astronomy, University of Potsdam, 14476 Potsdam-Golm, Germany}\affiliation{DESY, Platanenallee 6, 15738 Zeuthen, Germany}
\author{J.~H.~Buckley} 
\affiliation{Department of Physics, Washington University, St. Louis, MO 63130, USA}
\author{A.~J.~Chromey} 
\affiliation{Department of Physics and Astronomy, Iowa State University, Ames, IA 50011, USA}
\author{M.~K.~Daniel} 
\affiliation{Fred Lawrence Whipple Observatory, Harvard-Smithsonian Center for Astrophysics, Amado, AZ 85645, USA}
\author{A.~Falcone} 
\affiliation{Department of Astronomy and Astrophysics, 525 Davey Lab, Pennsylvania State University, University Park, PA 16802, USA}
\author{J.~P.~Finley} 
\affiliation{Department of Physics and Astronomy, Purdue University, West Lafayette, IN 47907, USA}
\author{L.~Fortson} 
\affiliation{School of Physics and Astronomy, University of Minnesota, Minneapolis, MN 55455, USA}
\author{A.~Furniss} 
\affiliation{Department of Physics, California State University - East Bay, Hayward, CA 94542, USA}
\author{A.~Gent} 
\affiliation{School of Physics and Center for Relativistic Astrophysics, Georgia Institute of Technology, 837 State Street NW, Atlanta, GA 30332-0430}
\author{G.~H.~Gillanders} 
\affiliation{School of Physics, National University of Ireland Galway, University Road, Galway, Ireland}
\author{D.~Hanna} 
\affiliation{Physics Department, McGill University, Montreal, QC H3A 2T8, Canada}
\author{T.~Hassan} 
\affiliation{DESY, Platanenallee 6, 15738 Zeuthen, Germany}
\author{O.~Hervet} 
\affiliation{Santa Cruz Institute for Particle Physics and Department of Physics, University of California, Santa Cruz, CA 95064, USA}
\author{J.~Holder} 
\affiliation{Department of Physics and Astronomy and the Bartol Research Institute, University of Delaware, Newark, DE 19716, USA}
\author{G.~Hughes} 
\affiliation{Fred Lawrence Whipple Observatory, Harvard-Smithsonian Center for Astrophysics, Amado, AZ 85645, USA}
\author{T.~B.~Humensky} 
\affiliation{Physics Department, Columbia University, New York, NY 10027, USA}
\author{P.~Kaaret} 
\affiliation{Department of Physics and Astronomy, University of Iowa, Van Allen Hall, Iowa City, IA 52242, USA}
\author{P.~Kar} 
\affiliation{Department of Physics and Astronomy, University of Utah, Salt Lake City, UT 84112, USA}
\author{M.~Kertzman} 
\affiliation{Department of Physics and Astronomy, DePauw University, Greencastle, IN 46135-0037, USA}
\author{D.~Kieda} 
\affiliation{Department of Physics and Astronomy, University of Utah, Salt Lake City, UT 84112, USA}
\author{M.~Krause} 
\affiliation{DESY, Platanenallee 6, 15738 Zeuthen, Germany}
\author{F.~Krennrich} 
\affiliation{Department of Physics and Astronomy, Iowa State University, Ames, IA 50011, USA}
\author{S.~Kumar} 
\affiliation{Physics Department, McGill University, Montreal, QC H3A 2T8, Canada}\affiliation{Department of Physics and Astronomy and the Bartol Research Institute, University of Delaware, Newark, DE 19716, USA}
\author{M.~J.~Lang} 
\affiliation{School of Physics, National University of Ireland Galway, University Road, Galway, Ireland}
\author{T.~T.~Y.~Lin} 
\affiliation{Physics Department, McGill University, Montreal, QC H3A 2T8, Canada}
\author{G.~Maier} 
\affiliation{DESY, Platanenallee 6, 15738 Zeuthen, Germany}
\author{P.~Moriarty} 
\affiliation{School of Physics, National University of Ireland Galway, University Road, Galway, Ireland}
\author{R.~Mukherjee} 
\affiliation{Department of Physics and Astronomy, Barnard College, Columbia University, NY 10027, USA}
\author{S.~O'Brien} 
\affiliation{School of Physics, University College Dublin, Belfield, Dublin 4, Ireland}
\author{R.~A.~Ong} 
\affiliation{Department of Physics and Astronomy, University of California, Los Angeles, CA 90095, USA}
\author{A.~N.~Otte} 
\affiliation{School of Physics and Center for Relativistic Astrophysics, Georgia Institute of Technology, 837 State Street NW, Atlanta, GA 30332-0430}
\author{N.~Park} 
\affiliation{WIPAC and Department of Physics, University of Wisconsin-Madison, Madison WI, USA}
\author{A.~Petrashyk} 
\affiliation{Physics Department, Columbia University, New York, NY 10027, USA}
\author{M.~Pohl} 
\affiliation{Institute of Physics and Astronomy, University of Potsdam, 14476 Potsdam-Golm, Germany}\affiliation{DESY, Platanenallee 6, 15738 Zeuthen, Germany}
\author{E.~Pueschel} 
\affiliation{DESY, Platanenallee 6, 15738 Zeuthen, Germany}
\author{J.~Quinn} 
\affiliation{School of Physics, University College Dublin, Belfield, Dublin 4, Ireland}
\author{K.~Ragan} 
\affiliation{Physics Department, McGill University, Montreal, QC H3A 2T8, Canada}
\author{G.~T.~Richards} 
\affiliation{School of Physics and Center for Relativistic Astrophysics, Georgia Institute of Technology, 837 State Street NW, Atlanta, GA 30332-0430}\affiliation{Department of Physics and Astronomy and the Bartol Research Institute, University of Delaware, Newark, DE 19716, USA}
\author{E.~Roache} 
\affiliation{Fred Lawrence Whipple Observatory, Harvard-Smithsonian Center for Astrophysics, Amado, AZ 85645, USA}
\author{I.~Sadeh} 
\affiliation{DESY, Platanenallee 6, 15738 Zeuthen, Germany}
\author{M.~Santander} 
\affiliation{Department of Physics and Astronomy, University of Alabama, Tuscaloosa, AL 35487, USA}
\author{S.~Schlenstedt} 
\affiliation{DESY, Platanenallee 6, 15738 Zeuthen, Germany}
\author{G.~H.~Sembroski} 
\affiliation{Department of Physics and Astronomy, Purdue University, West Lafayette, IN 47907, USA}
\author{I.~Sushch} 
\affiliation{DESY, Platanenallee 6, 15738 Zeuthen, Germany}
\author{J.~Tyler} 
\affiliation{Physics Department, McGill University, Montreal, QC H3A 2T8, Canada}
\author{V.~V.~Vassiliev} 
\affiliation{Department of Physics and Astronomy, University of California, Los Angeles, CA 90095, USA}
\author{S.~P.~Wakely} 
\affiliation{Enrico Fermi Institute, University of Chicago, Chicago, IL 60637, USA}
\author{A.~Weinstein} 
\affiliation{Department of Physics and Astronomy, Iowa State University, Ames, IA 50011, USA}
\author{R.~M.~Wells} 
\affiliation{Department of Physics and Astronomy, Iowa State University, Ames, IA 50011, USA}
\author{P.~Wilcox} 
\affiliation{Department of Physics and Astronomy, University of Iowa, Van Allen Hall, Iowa City, IA 52242, USA}
\author{A.~Wilhelm} 
\affiliation{Institute of Physics and Astronomy, University of Potsdam, 14476 Potsdam-Golm, Germany}\affiliation{DESY, Platanenallee 6, 15738 Zeuthen, Germany}
\author{D.~A.~Williams} 
\affiliation{Santa Cruz Institute for Particle Physics and Department of Physics, University of California, Santa Cruz, CA 95064, USA}
\author{T.~J~Williamson} 
\affiliation{Department of Physics and Astronomy and the Bartol Research Institute, University of Delaware, Newark, DE 19716, USA}
\author{B.~Zitzer} 
\affiliation{Physics Department, McGill University, Montreal, QC H3A 2T8, Canada}
\collaboration{(VERITAS collaboration)}

\author{V.~A.~Acciari} 
\affiliation{Inst. de Astrof\'isica de Canarias, E-38200 La Laguna, and Universidad de La Laguna, Dpto. Astrof\'isica, E-38206 La Laguna, Tenerife, Spain}
\author{S.~Ansoldi} 
\affiliation{Universit\`a di Udine, and INFN Trieste, I-33100 Udine, Italy}\affiliation{Japanese MAGIC Consortium: ICRR, The University of Tokyo, 277-8582 Chiba, Japan; Department of Physics, Kyoto University, 606-8502 Kyoto, Japan; Tokai University, 259-1292 Kanagawa, Japan; RIKEN, 351-0198 Saitama, Japan}
\author{L.~A.~Antonelli} 
\affiliation{National Institute for Astrophysics (INAF), I-00136 Rome, Italy}
\author{A.~Arbet Engels} 
\affiliation{ETH Zurich, CH-8093 Zurich, Switzerland}
\author{D.~Baack} 
\affiliation{Technische Universit\"at Dortmund, D-44221 Dortmund, Germany}
\author{A.~Babi\'c} 
\affiliation{Croatian MAGIC Consortium: University of Rijeka, 51000 Rijeka; University of Split - FESB, 21000 Split; University of Zagreb - FER, 10000 Zagreb; University of Osijek, 31000 Osijek; Rudjer Boskovic Institute, 10000 Zagreb, Croatia}
\author{B.~Banerjee} 
\affiliation{Saha Institute of Nuclear Physics, HBNI, 1/AF Bidhannagar, Salt Lake, Sector-1, Kolkata 700064, India}
\author{U.~Barres de Almeida} 
\affiliation{Centro Brasileiro de Pesquisas F\'isicas (CBPF), 22290-180 URCA, Rio de Janeiro (RJ), Brasil}
\author{J.~A.~Barrio} 
\affiliation{Unidad de Part\'iculas y Cosmolog\'ia (UPARCOS), Universidad Complutense, E-28040 Madrid, Spain}
\author{J.~Becerra Gonz\'alez} 
\affiliation{Inst. de Astrof\'isica de Canarias, E-38200 La Laguna, and Universidad de La Laguna, Dpto. Astrof\'isica, E-38206 La Laguna, Tenerife, Spain}
% ...
\collaboration{(MAGIC collaboration)}

\correspondingauthor{Ralph Bird, Javier Herrera, Alicia L\'opez-Oramas and Tyler Williamson}
\email{ralphbird@astro.ucla.edu, jaherllo@iac.es, aloramas@iac.es, tjwilli@udel.edu}

\begin{abstract}
We report on observations of the pulsar / Be star binary system \psrbe\ in the energy range between $100\U{GeV}$ and $20\U{TeV}$ with the VERITAS and MAGIC imaging atmospheric Cherenkov telescope arrays. The binary orbit has a period of approximately 50 years, with the most recent periastron occurring on 2017 November 13. Our observations span from 18 months prior to periastron to one month after. A new, point-like, gamma-ray source is detected, coincident with the location of \psrbe. The gamma-ray light curve and spectrum are well-characterized over the periastron passage. The flux is variable over at least an order of magnitude, peaking at periastron, thus providing a firm association of the TeV source with the pulsar / Be star system. Observations prior to periastron show a cutoff in the spectrum at an energy around $0.5\U{TeV}$.  This result adds a new member to the small population of known TeV binaries, and it identifies only the second source of this class in which the nature and properties of the compact object are firmly established.  

We compare the gamma-ray results with the light curve measured with the X-ray Telescope (XRT) on board the Neil Gehrels \textit{Swift} Observatory and with the predictions of recent theoretical models of the system. We conclude that significant revision of the models is required to explain the details of the emission we have observed, and we discuss the relationship between the binary system and the overlapping steady extended source, \tevj. 
\end{abstract}


\section{Introduction} \label{sec:intro}

TeV gamma-ray emitting binary systems are extremely rare objects, likely corresponding to a relatively brief period in the evolution of some massive star binaries \citep{2017AnA...608A..59D}. They consist of a neutron star or black hole in a binary orbit with a massive star, and their high-energy emission provides a unique opportunity to study relativistic particle acceleration in a continuously changing physical environment. Of particular interest are systems where the compact object is a pulsar, since the pulsed emission firmly identifies the nature of the compact object, and provides an accurate determination of the orbital parameters, the available energy budget, and the likely acceleration mechanism. Prior to this work, only one TeV binary with a known pulsar had been detected: the pulsar / Be-star binary system PSR\,B1259-63 / LS\,2883 \citep{2005A&A...442....1A}. In this paper, we present the discovery of a second member of this class, PSR\,J2032+4127 / MT91\,213.

Pulsed emission, with a period of $P=143\U{ms}$, was first detected from \psr\ in a blind search of \textit{Fermi}-LAT gamma-ray data \citep{2009Sci...325..840A} and was subsequently detected in radio observations with the Green Bank Telescope \citep{2009ApJ...705....1C}. These observations revealed dramatic changes in the pulsar spin-down rate, an effect most easily explained by Doppler shift due to the pulsar's motion in a long-period binary system \citep{2015MNRAS.451..581L}. The pulsar's companion was identified as a B0Ve star, MT91\,213, which has a mass of around 15$\U{M_\odot}$ and a circumstellar disk which varies in radius by more than a factor of two, from $0.2\U{AU}$ to $0.5\U{AU}$ \citep{2017MNRAS.464.1211H}. The pulsar spin-down luminosity ($\dot{\mathrm{E}}$) is $1.7\times 10^{35}\U{erg}\UU{s}{-1} $, with a characteristic age of 180\U{kyr}, and the system lies at a distance of 1.4--1.7$\U{kpc}$, in the Cyg OB2 stellar association. Further observations refined the orbital parameters, yielding a binary period of 45--50 years, an eccentricity between 0.94 and 0.99, and a longitude of periastron between 21\arcdeg and 52\arcdeg. Periastron occurred on 2017 November 13  with a separation between \psr\ and \be\ of approximately 1~AU \citep{2017MNRAS.464.1211H, 2017ATel10920....1C}. 

Significant X-ray brightening from the direction of \psr\ was first detected by \citet{2017MNRAS.464.1211H}, with the X-ray flux increasing by a factor of twenty relative to 2010 measurements. \citet{2017ApJ...843...85L} used data from \textit{Chandra}, \textit{Swift}-XRT, \textit{NuSTAR}, and \textit{XMM-Newton} to conduct a detailed study of the long-term light curve, finding variability on timescales of weeks on top of the long-term increasing trend, which they attributed to clumps in the stellar wind. The structure of the stellar wind was further explored by \citet{2018MNRAS.474L..22P}, who used \textit{Swift}-XRT observations to map the circumstellar environment of the Be star. Recently, \citet{2018ApJ...857..123L} presented detailed \textit{Swift}, \textit{Fermi}-LAT and radio observations of \psr\ over the 2017 periastron period. They report strong variability in the X-ray flux, but no variability in the GeV gamma-ray flux, likely because this is masked by magnetospheric emission from the pulsar.

\psr\ lies at the edge of the steady, extended, very-high-energy (VHE, $E>100\U{GeV}$) gamma-ray source, \tevj. This object was the first VHE source to be serendipitously discovered, by HEGRA \citep{2002A&A...393L..37A}, and was not associated with any counterpart at other wavelengths. Subsequent observations by HEGRA and MAGIC revealed an extended object, with a width of approximately 6\arcmin\ and a hard power-law spectrum ($\Gamma\sim2.0$) \citep{2005A&A...431..197A, 2008ApJ...675L..25A}. VERITAS observations have shown that the extended emission is asymmetric and coincident with a void in radio emission \citep{2014ApJ...783...16A}. Prior to the discovery of its binary nature, an association of \tevj\ with the pulsar wind nebula (PWN) of PSR\,J2032+4127 seemed the most likely origin for the VHE source.

Thus far, only a handful of VHE gamma-ray emitting binary systems have been detected, of which only PSR\,B1259-63 / LS\,2883 has an identified compact object: a pulsar in a highly elliptical ($e=0.87$) orbit with a period of 3.4~years  and a periastron separation of about $1\U{AU}$ \citep{1992ApJ...387L..37J,2011ApJ...732L..11N}.
The unpulsed radio, X-ray and VHE gamma-ray fluxes show complex light curves, with the majority of the emission occurring close to periastron in two distinct peaks, likely related to the pulsar passage through the circumstellar decretion disk. High-energy (HE, $0.1\U{GeV}<E<100\U{GeV}$) emission, conversely, is generally weak around periastron, followed by intense flaring episodes which occur typically more than 30, and as long as 70, days after periastron \citep{2011ApJ...736L..11A,2018ApJ...863...27J}. 

In this paper, we report on the results of extensive VHE gamma-ray observations of the 2017 periastron passage of the \psrbe\ system using VERITAS and MAGIC, as well as presenting contemporaneous X-ray observations with \textit{Swift}-XRT. 

\section{Observations and Analysis} \label{sec:observations}

VHE gamma-ray observations of \psrbe\ were conducted by the MAGIC and VERITAS imaging atmospheric Cherenkov telescope arrays, which are sensitive to astrophysical gamma rays above $50\U{GeV}$. MAGIC consists of two 17\,m-diameter telescopes, located at the observatory of El Roque de Los Muchachos on the island of La Palma, Spain, whereas VERITAS is an array of four 12\,m-diameter telescopes at the Fred Lawrence Whipple Observatory near Tucson, Arizona. The observatories and their capabilities are described in \citet{Aleksic2016} (MAGIC) and \citet{2015ICRC...34..771P} (VERITAS), and references therein.

In this paper, we present the results of 181.3 hours of observations with VERITAS (51.6 hours of archival data taken before 2016, 30.1 hours between 2016 September and 2017 June, 99.6 hours between 2017 September and December) and 87.9 hours of observations with MAGIC (53.7 hours between 2016 May and September and 34.2 hours between 2017 June and December). Observations were conducted in ``wobble'' mode, with the source location offset from the center of the field-of-view, allowing simultaneous evaluation of the background \citep{Fomin94}. The data were analyzed using standard tools \citep{Zanin13, 2017arXiv170804048M}, in which Cherenkov images are first calibrated, cleaned and parameterized \citep{Hillas85}, then used to reconstruct the energy and arrival direction of the incident gamma ray, and to reject the majority of the cosmic ray background \citep{magic:RF, 2006APh....25..380K}. 

A total of 186 \textit{Swift}-XRT \citep{2005SSRv..120..165B} observations were taken between 2008 June 16 and 2018 April 15, equating to 136.4 hours of live time. The data were collected in photon-counting mode and analyzed using the HEAsoft analysis package version 6.24\footnote{\url{https://heasarc.nasa.gov/lheasoft/}}. The background was estimated from five regions equidistant from \psr, and the flux was calculated using XSPEC.

\section{Variability, Morphology and Spectrum} \label{sec:results}

A new, spatially-unresolved, time-varying VHE gamma-ray source was detected at a position compatible with \psrbe. The source is named \vername\ and \magname\ in the VERITAS and MAGIC source catalogs, respectively. It is also spatially coincident with \tevj, the previously detected extended VHE source, but offset from the centroid of the extended emission by approximately $10\arcmin$.

\begin{figure*}[ht!]
\gridline{\leftfig{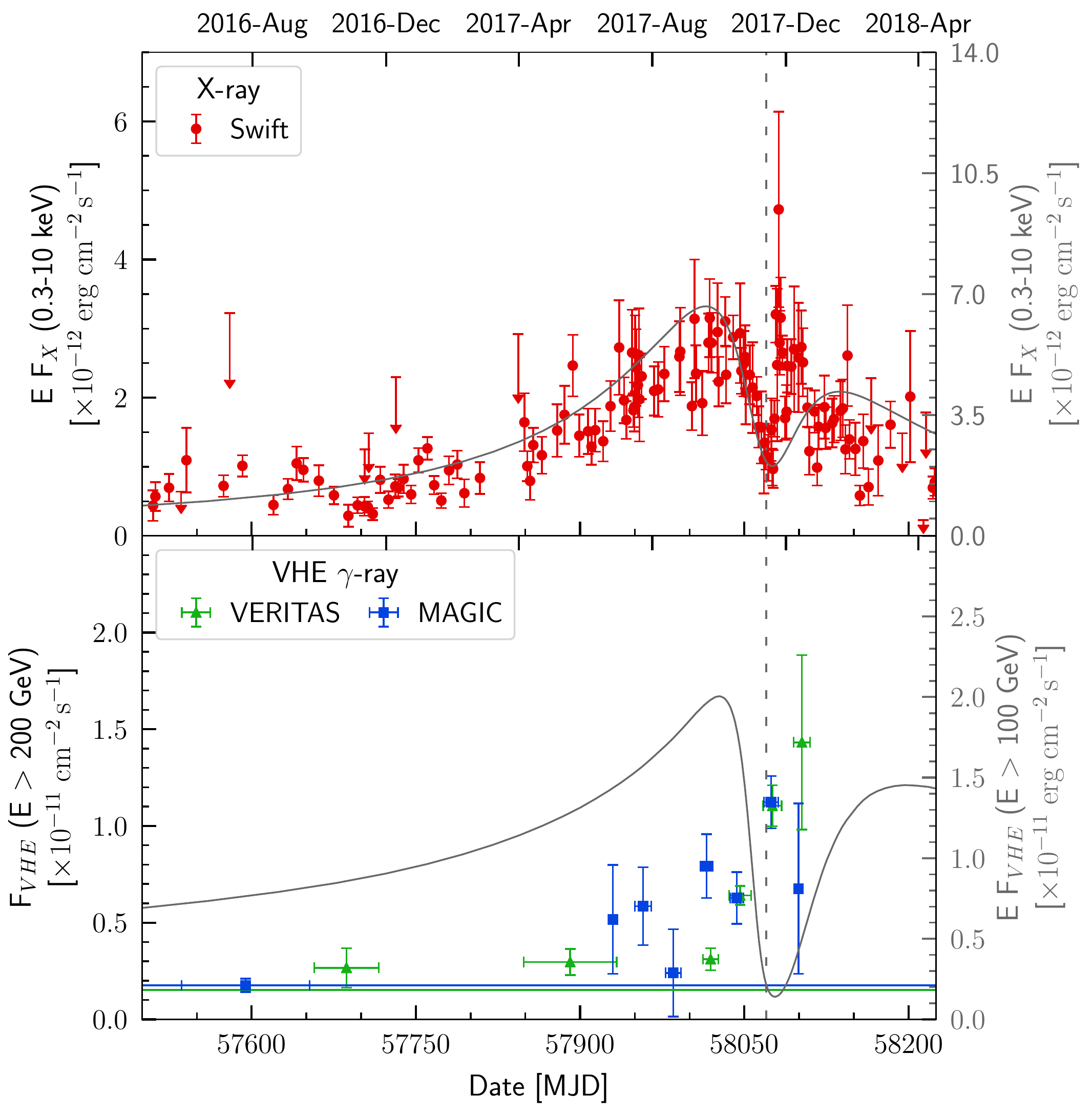}{\columnwidth}{(a) Full Dataset}
\rightfig{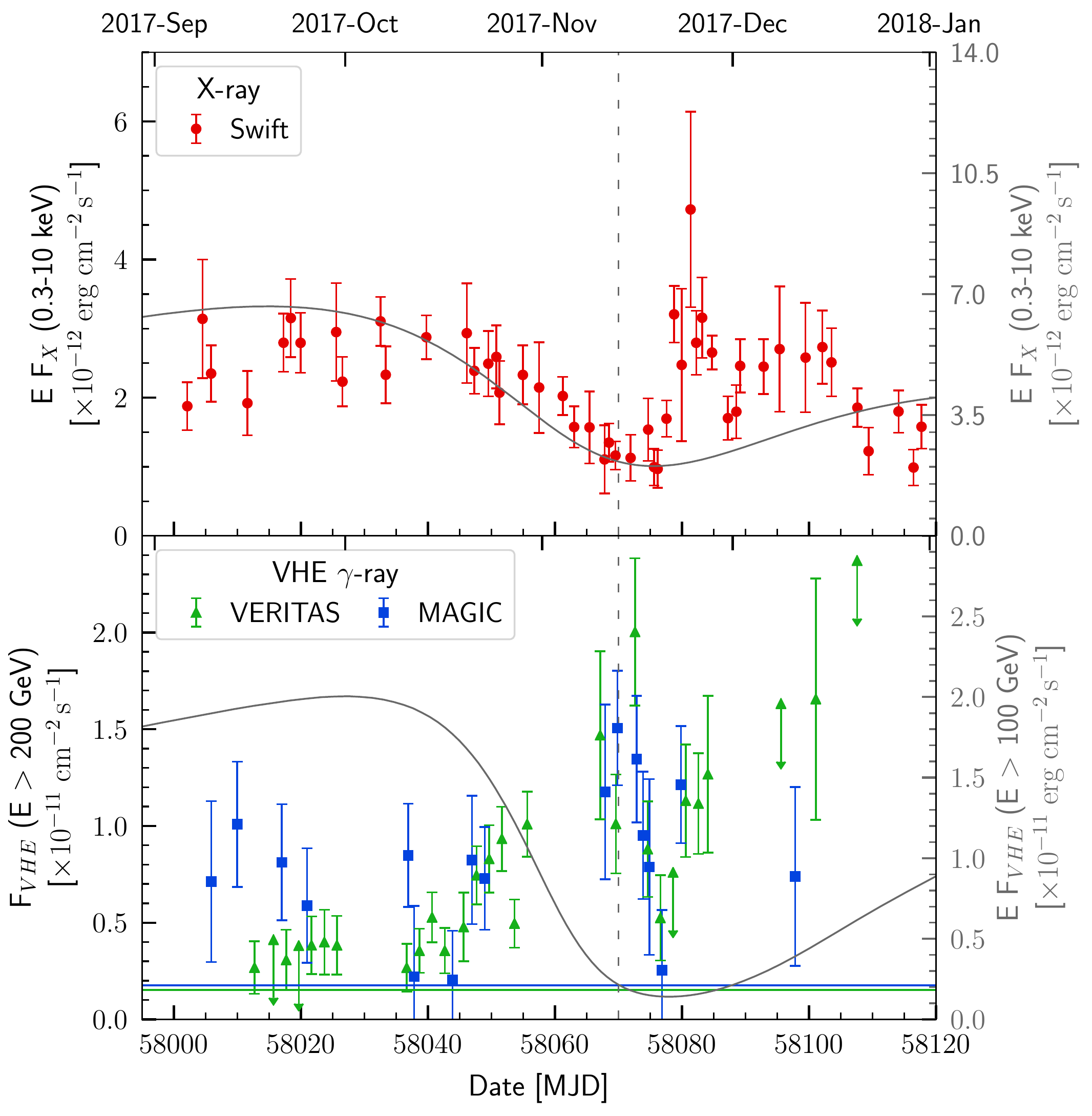}{\columnwidth}{(b) Periastron}}
\caption{Upper panels (left axes) show the 0.3--10.0~keV background-subtracted \textit{Swift}-XRT energy-flux light curve (red circles) of \psrbe. For clarity, observations with exposures less than 1.4 ks are excluded from the plot. Lower panels show the $>200\U{GeV}$ photon-flux light curves from VERITAS (green triangles) and MAGIC (blue squares). The left plot shows the full light curve, while the right plot shows only the months around periastron. The horizontal solid lines indicate the average flux prior to 2017 for the respective experiments.  The solid gray lines (right axes) are the energy-flux light curve predictions from \cite{2018ApJ...857..123L} for X-rays and updated predictions from \cite{2017ApJ...836..241T} using the  parameters from \cite{2018ApJ...857..123L} (Takata, private communication) for VHE gamma rays. Both models assume an inclination angle of $60\arcdeg$. The vertical gray dashed line indicates periastron.
\label{fig:LC}}
\end{figure*}

The complete X-ray and gamma-ray light curves are shown in \autoref{fig:LC}. 
Following the initial detection of \psrbe\ in 2017 September by VERITAS and MAGIC \citep{2017ATel10810....1V} with a flux exceeding the baseline flux from \tevj, gamma-ray emission was observed to increase up to the time of periastron (2017 November 13; MJD 58070), reaching a factor of ten higher than the baseline. Approximately one week after periastron the flux sharply decreased, to a level compatible with the baseline emission, before recovering to the periastron level a few days later. Further observations were conducted after periastron, but the combination of low source elevation angle, poor weather conditions and rather brief exposure, resulted in a relatively poor flux measurement. In total, during the 2017 fall observations (MJD 57997--58110), VERITAS detected \psrbe\ with a significance of 21.5 standard deviations ($\sigma$) and MAGIC with a significance of $19.5\U{\sigma}$. 

\begin{figure*}[ht!]
\centering
\gridline{\leftfig{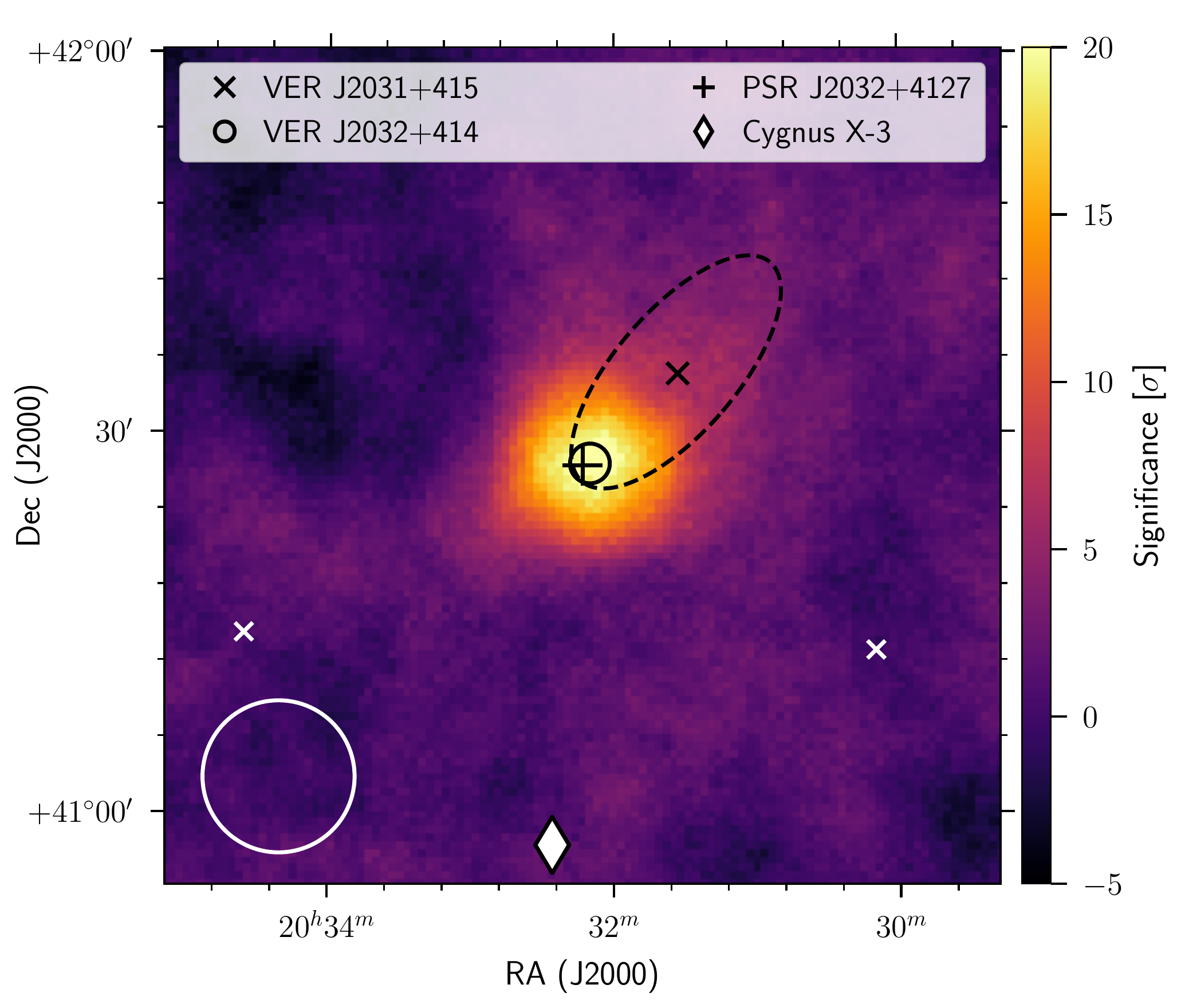}{\columnwidth}{(a) VERITAS 2017 fall sky map}
\rightfig{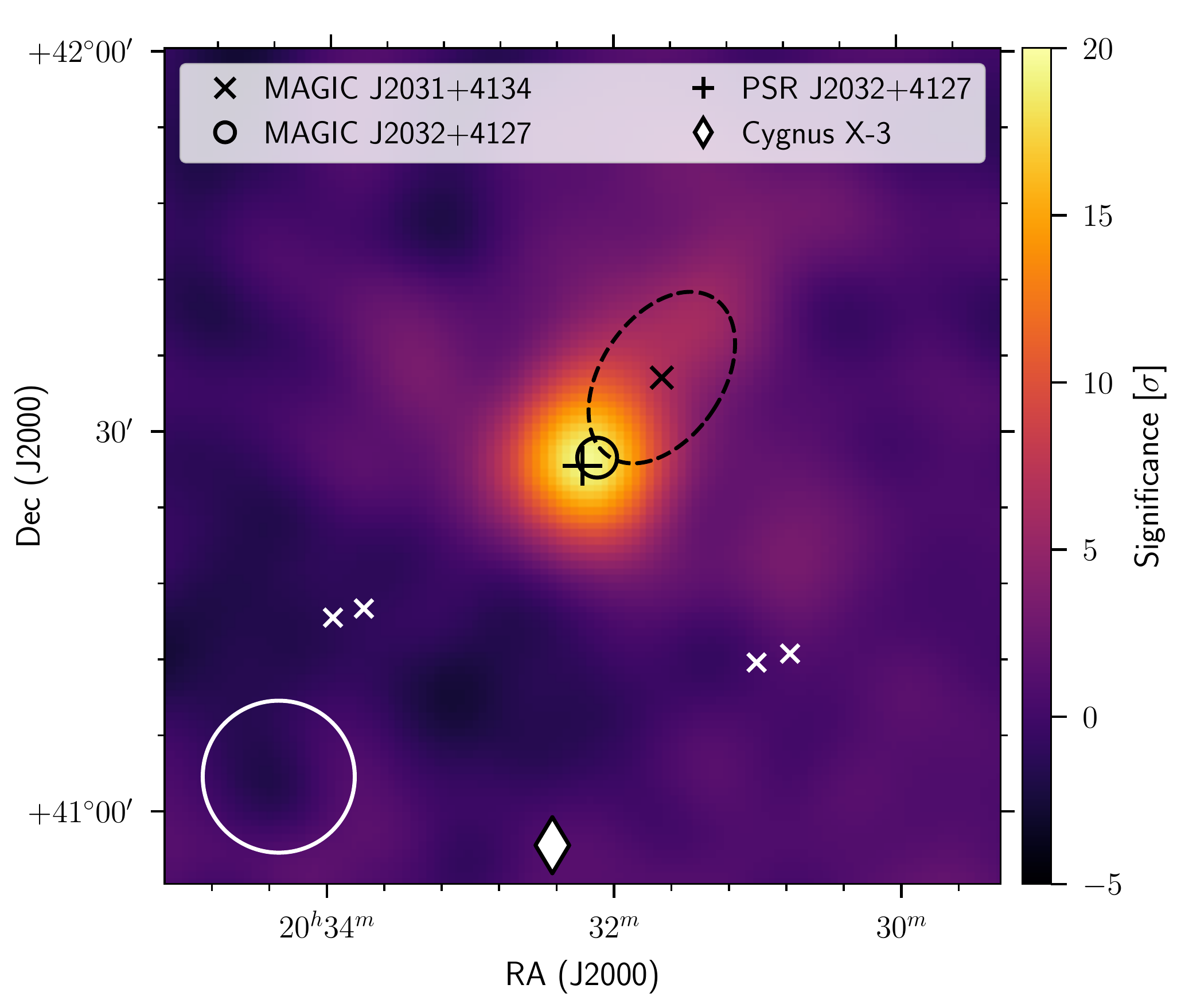}{\columnwidth}{(b) MAGIC 2017 fall sky map}}
\caption{Significance sky maps of the region around \psrbe\ showing both the VERITAS (left) and MAGIC (right) results for observations during 2017 fall.
The position of \psrbe\ is shown as a black ``{\bf+}'', the centroid of the gamma-ray emission as a black ``{\bf$\circ$}'', the position and extension for the respective telescope's measurements of \tevj\ are shown as a black ``{\bf$\times$}'' and a dashed line, and the position of Cygnus X-3 is shown with a white diamond.  The white circle in the lower left hand corner is of radius 0\fdg1, the approximate point spread function for these measurements at $1\U{TeV}$. The wobble positions are shown as white ``{\bf$\times$}''. 
\label{fig:maps}}
\end{figure*}

\autoref{fig:maps} shows skymaps for the complete fall 2017 VHE datasets, revealing overlapping emission from \tevj\ and \psrbe.  For both the VERITAS and MAGIC data we fit the gamma-ray excess maps with a two-component model, consisting of a bivariate Gaussian function to represent the extended source and a symmetrical Gaussian function to model the unresolved emission at the location of the binary. The parameters of the extended source model, indicated by the dashed ellipses in \autoref{fig:maps}, were constrained to match those measured prior to the appearance of the binary. For MAGIC, the extended source has a semi-major axis of ${0\fdg125}\pm{0\fdg01}$ and a semi-minor axis of ${0\fdg08}\pm{0\fdg01}$, centered on R.A.=$20^\mathrm{h} 31^\mathrm{m} 39.7^\mathrm{s} \pm 2^\mathrm{s}$, Dec=$41\arcdeg 34\arcmin 23\arcsec \pm 20\arcsec$, at an angle of $34\arcdeg \pm 2\arcdeg$ east of north. For VERITAS, the extended source parameters are those reported in \citet{2018ApJ...861..134A}: semi-major and semi-minor axes of ${0\fdg19}\pm{0\fdg02}_{stat}\pm{0\fdg01}_{sys}$ and ${0\fdg08}\pm{0\fdg01}_{stat}\pm{0\fdg03}_{sys}$, centered on R.A.=$20^\mathrm{h} 31^\mathrm{m} 33^\mathrm{s}\pm 2^\mathrm{s}_\mathrm{stat}\pm 2^\mathrm{s}_\mathrm{sys}$, Dec=$41\arcdeg 34\arcmin 38\arcsec \pm36\arcsec_\mathrm{stat} \pm36\arcsec_\mathrm{sys}$, with an orientation of $41\arcdeg \pm{4\arcdeg}_{stat}\pm{1\arcdeg}_{sys}$ east of north.  The centroid of the unresolved component is measured to be at R.A.=$20^\mathrm{h} 32^\mathrm{m} 10^\mathrm{s} \pm 2^\mathrm{s}_\mathrm{stat} \pm 2^\mathrm{s}_\mathrm{sys}$, Dec=$41\arcdeg 27\arcmin 34\arcsec \pm 16\arcsec_\mathrm{stat} \pm 26\arcsec_\mathrm{sys}$ for VERITAS, and R.A.=$20^\mathrm{h} 32^\mathrm{m} 7^\mathrm{s} \pm 2_\mathrm{stat}^\mathrm{s}$, Dec=$41\arcdeg 28\arcmin 4\arcsec \pm 20\arcsec_\mathrm{stat}$ for MAGIC, which are consistent, within the measured uncertainties, with the location of \psrbe\ ($20^\mathrm{h} 32^\mathrm{m} 13.12^\mathrm{s} \pm 0.02^\mathrm{s} +41\arcdeg 27\arcmin 24.34\arcsec \pm 0.03\arcsec$; \cite{2018yCat.1345....0G}).

\begin{figure*}[ht!]
\centering
\gridline{\leftfig{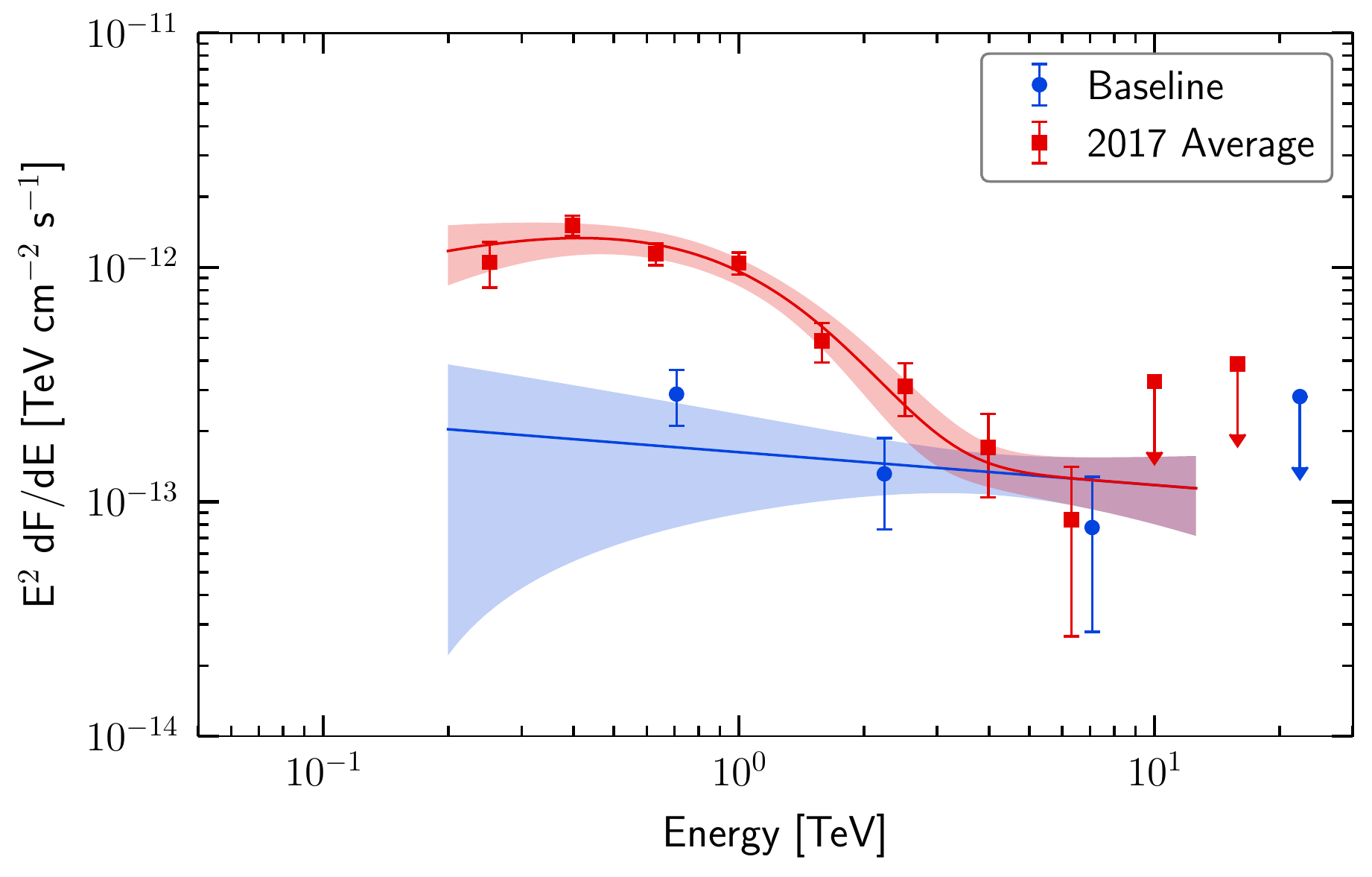}{\columnwidth}{(a) VERITAS 2017 fall average\label{fig:VERAverage}}
\rightfig{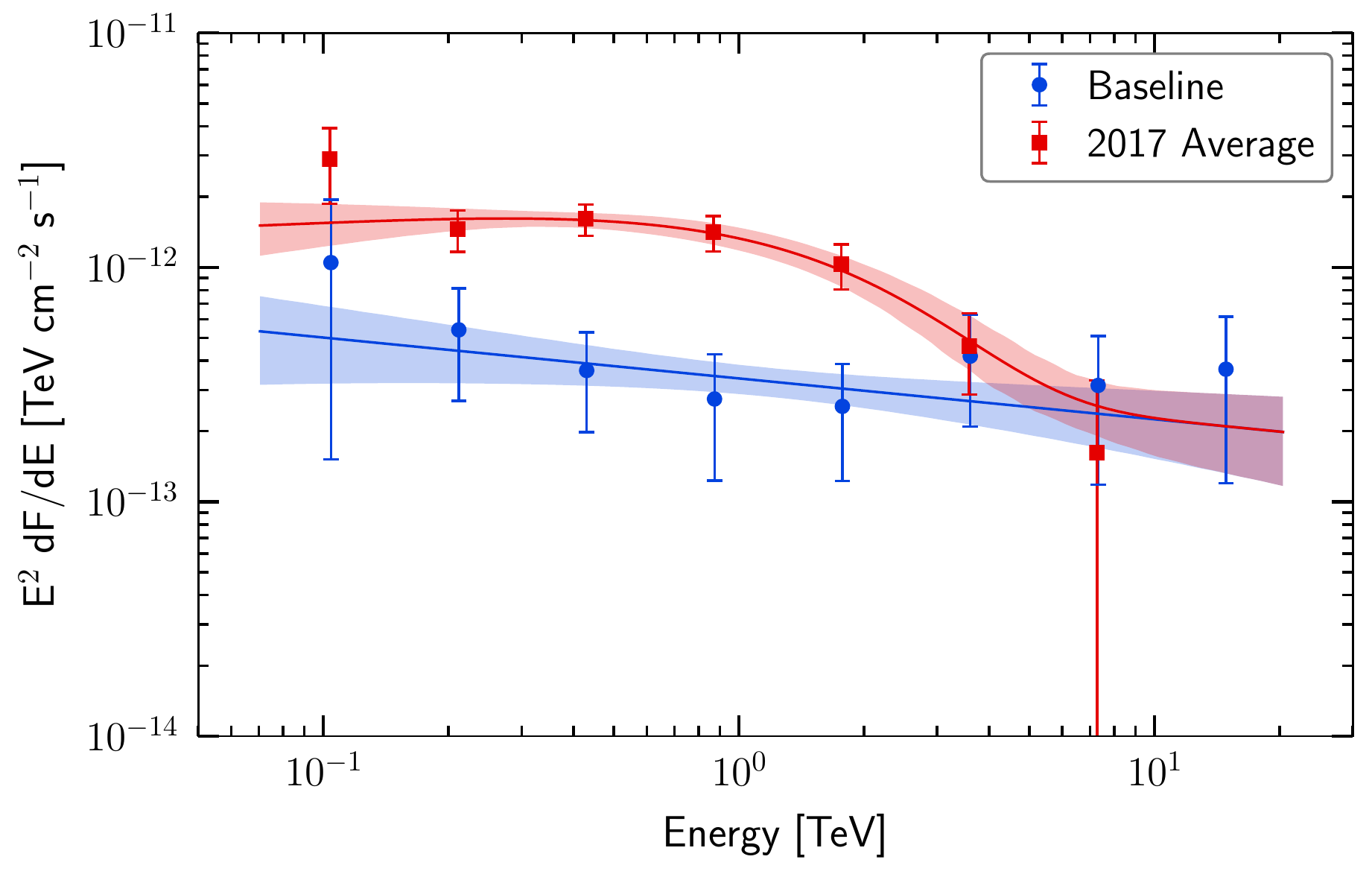}{\columnwidth}{(b) MAGIC 2017 fall average}}
\gridline{\leftfig{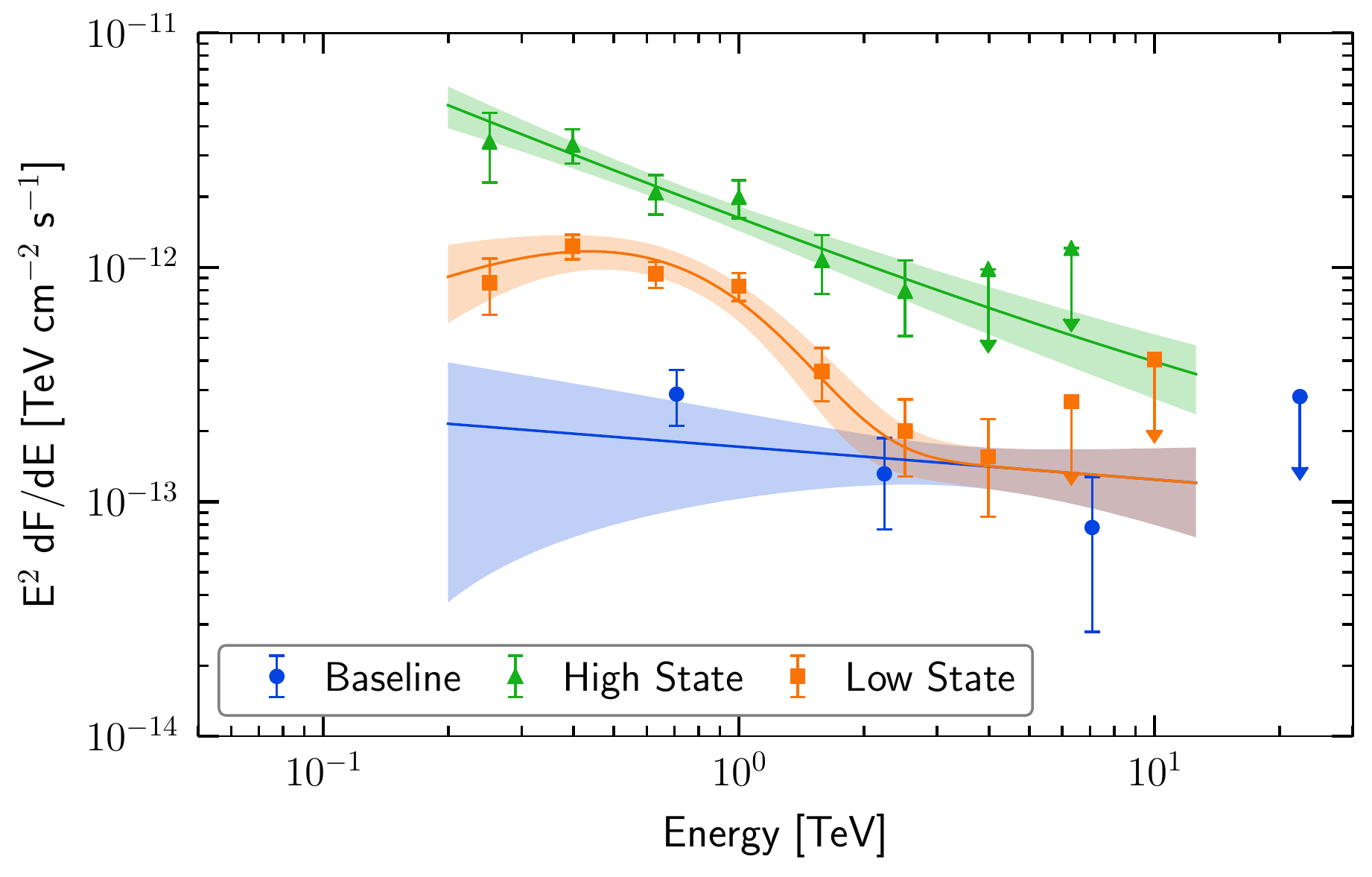}{\columnwidth}{(c) VERITAS high \& low states}
\rightfig{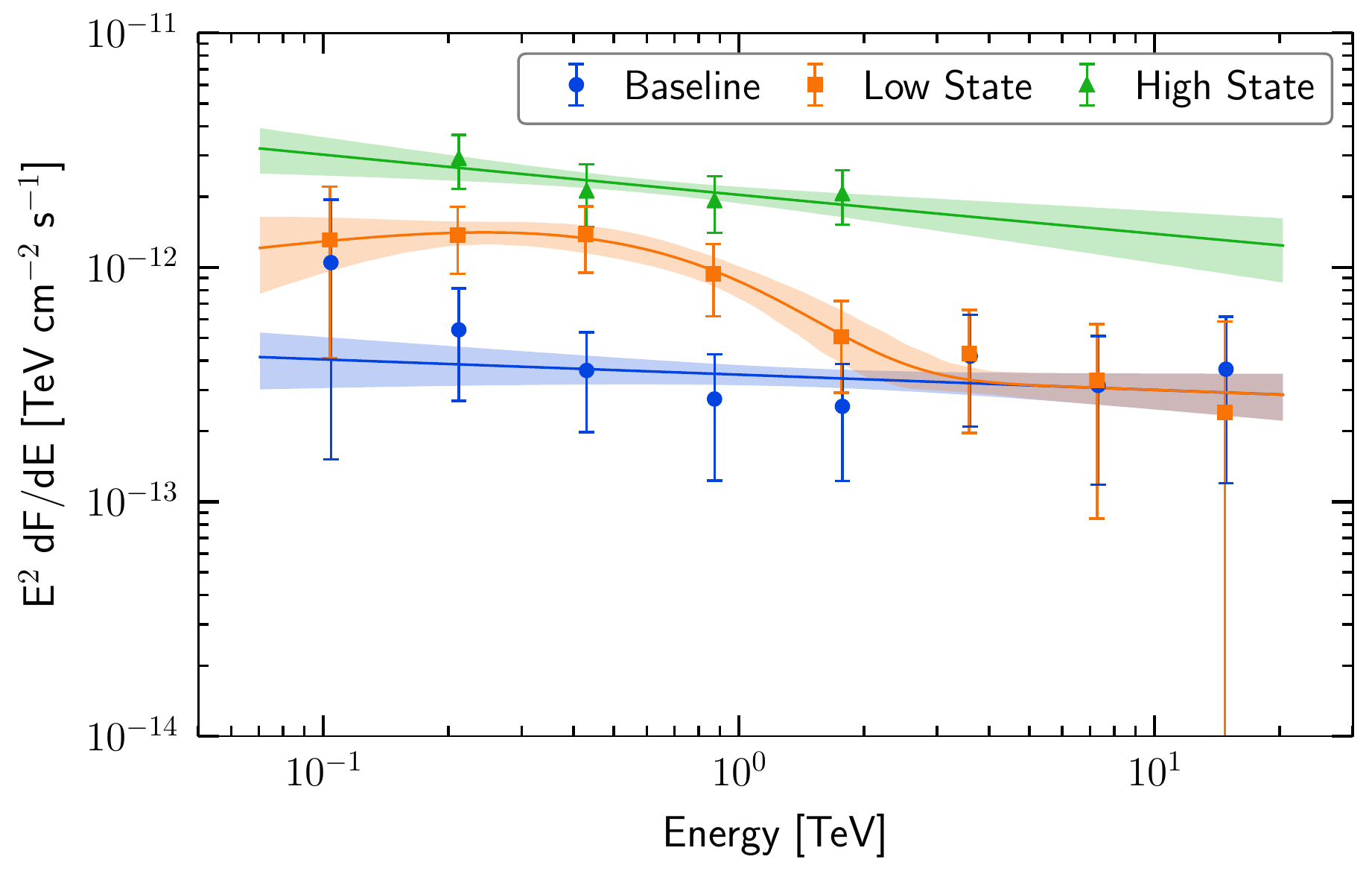}{\columnwidth}{(d) MAGIC high \& low states}}
\caption{Spectral energy distributions for \psrbe\ and \tevj\ from VERITAS (left) and MAGIC (right).  The blue butterflies are the spectral fits to \tevj. The red butterflies in the upper plots are fits to the 2017 fall data: the sum of a power-law fit to \tevj\ and a cutoff power-law fit to \psrbe.  In the bottom plots, orange is the fit to the low-state data (\psrbe\ is fit with a cutoff) while green represents the high-state data (\psrbe\ is fit with a power law).  The fit parameters are given in \autoref{tab:spectralFits} and the time periods are defined in the text.  \label{fig:spectra}}
\end{figure*}

The source spectrum (\autoref{fig:spectra}) is also formed of two emission components: steady, baseline emission from the extended source \tevj\ and variable emission associated with the binary system. Prior to 2017, only the baseline emission component was present, while the 2017 data include contributions from both the baseline and the variable binary. We performed a global spectral fit to the complete dataset, in which the pre-2017 observations were fit with a pure power-law for the baseline, and the 2017 data were fit with the same power-law, plus an additional component for the binary emission. Two models were tested for the binary emission: a pure power law and a power law with an exponential cutoff. The VERITAS data favor the cutoff model over the power law for the binary emission, with an F-test probability of 0.997 and a cutoff energy of $0.57\pm0.20\U{TeV}$. MAGIC observations also favor an exponential cutoff, with a probability of 0.993 and a cutoff energy of $1.40\pm0.97\U{TeV}$. Full details of the fit parameters are given in \autoref{tab:spectralFits}. We note that the only other gamma-ray binary to display a spectral cutoff in the VHE regime is LS\,5039, with a cutoff at $8.7\pm2.0\U{TeV}$ in the VHE high state, close to inferior conjunction \citep{2006A&A...460..743A}.

The fit process was then repeated with the 2017 data broken up into two periods, to search for spectral variation with orbital phase and/or flux state of the binary system. We define a high state (MJD 58057--58074 and 58080--58110), which covers the periods around periastron where the flux above 0.2 TeV was greater than 1.0$\times10^{-11}$~cm$^{-2}$ s$^{-1}$ (approximately five times greater than the baseline flux from \tevj), and a low state, covering the 2017 observations prior to periastron (MJD 57928--58056).  We performed a global fit to the datasets, with the high and low states fit with the baseline power law plus either a pure power law or a power-law with an exponential cutoff. The VERITAS data favor a cutoff model in the low state, with an F-test probability of 0.999 and a cutoff value of $0.33\pm0.13\U{TeV}$. MAGIC observations also favor a low-state cutoff model, with a probability of 0.980 and a cutoff value of $0.58\pm0.33\U{TeV}$. For both observatories, the high-state data are well-fit by a pure power law and including a cutoff does not significantly change the quality of the fit.  

\begin{deluxetable*}{lccccccc}
\tablenum{1}
\tablecaption{VHE gamma-ray spectral fit results. Each group of rows shows the result of a simultaneous fit of both the baseline emission from the region prior to the appearance of the binary, modeled as a power law (PL), and the sum of this baseline with a new component from the binary, modeled as either a power law or a power law with an exponential cutoff (PLEC). These fits were performed across the data periods defined in \autoref{sec:results}. In each row, the parameters shown correspond to the model component listed in \textbf{bold}, where N$_0$ is the differential flux normalization (calculated at the de-correlation energy E$_0$), $\Gamma$ is the spectral index, and E$_\mathrm{C}$ is the cutoff energy for PLEC models. The $\chi^2$ and degrees of freedom (dof) are calculated from the joint fit across the given data. \label{tab:spectralFits}}
\tablehead{
\colhead{} &
\colhead{Period} &
\colhead{Model Components} &
\colhead{N$_0$} &
\colhead{E$_0$} &
\colhead{$\Gamma$} &
\colhead{E$_\mathrm{C}$} &
\colhead{$\chi ^2$/dof}\\
\colhead{} &
\colhead{} &
\colhead{} &
\colhead{[cm$^{-2}$ s$^{-1}$ TeV$^{-1}$]} &
\colhead{[TeV]} &
\colhead{} &
\colhead{[TeV]}&
\colhead{}
}
\startdata
\multirow{13}{2cm}{\centering VERITAS ($>220$ GeV)} & Pre-2017& \textbf{PL$_\mathrm{baseline}$} & (8.78 $\pm$ 2.56) $\times 10^{-15}$ & 3.47 & 2.14 $\pm$ 0.53 & - & \multirow{2}{*} {\writeChitworow{40.6/7}} \\
& Fall 2017& PL$_\mathrm{baseline}$ + \textbf{PL$_\mathrm{binary}$} & (1.53 $\pm$ 0.14) $\times 10^{-12}$ & 0.70 & 2.81 $\pm$ 0.09 & - & \\
\cline{2-8}
& Pre-2017& \textbf{PL$_\mathrm{baseline}$} & (7.62 $\pm$ 1.51) $\times 10^{-15}$ & 4.18 & 2.14 $\pm$ 0.29 & - & \multirow{2}{*}{\writeChitworow{8.6/6}} \\
& Fall 2017& PL$_\mathrm{baseline}$ + \textbf{PLEC$_\mathrm{binary}$} & (8.04 $\pm$ 3.37) $\times 10^{-12}$ & 0.64 & 1.26 $\pm$ 0.45 & 0.57 $\pm$ 0.20 & \\
\cline{2-8}
& Pre-2017& \textbf{PL$_\mathrm{baseline}$} & (4.65 $\pm$ 1.18) $\times 10^{-15}$ & 4.98 & 2.14 $\pm$ 0.85 & - & \multirow{3}{*}{\hspace{-.5pt}\writeChithreerow{26.8/10}} \\
&Low State& PL$_\mathrm{baseline}$ + \textbf{PL$_\mathrm{binary}$} & (8.12 $\pm$ 3.77) $\times 10^{-14}$ & 1.76 & 2.86 $\pm$ 0.11 & - & \\
& High State& PL$_\mathrm{baseline}$ + \textbf{PL$_\mathrm{binary}$} & (9.69 $\pm$ 1.75) $\times 10^{-13}$ & 1.17 & 2.72 $\pm$ 0.15 & - & \\
\cline{2-8}
& Pre-2017& \textbf{PL$_\mathrm{baseline}$} & (1.23 $\pm$ 0.24) $\times 10^{-14}$ & 3.43 & 2.14 $\pm$ 0.28 & - & \multirow{3}{*}{\writeChithreerow{7.9/9}} \\
& Low State& PL$_\mathrm{baseline}$ + \textbf{PLEC$_\mathrm{binary}$} & (1.63 $\pm$ 1.12) $\times 10^{-11}$ & 0.56 & 0.65 $\pm$ 0.75 & 0.33 $\pm$ 0.13 & \\
& High State& PL$_\mathrm{baseline}$ + \textbf{PL$_\mathrm{binary}$}& (1.45 $\pm$ 0.18) $\times 10^{-12}$ & 1.00 & 2.73 $\pm$ 0.15 & - & \\
\cline{2-8}
& Pre-2017& \textbf{PL$_\mathrm{baseline}$} & (1.26 $\pm$ 0.25) $\times 10^{-14}$ & 3.39 & 2.14 $\pm$ 0.28 & - & \multirow{3}{*}{\writeChithreerow{7.2/8}} \\
& Low State& PL$_\mathrm{baseline}$ + \textbf{PLEC$_\mathrm{binary}$} & (1.64 $\pm$ 1.12) $\times 10^{-11}$ & 0.56 & 0.65 $\pm$ 0.75 & 0.33 $\pm$ 0.13 & \\
& High State& PL$_\mathrm{baseline}$ + \textbf{PLEC$_\mathrm{binary}$} & (1.20 $\pm$ 0.41) $\times 10^{-11}$ & 0.51 & 2.37 $\pm$ 0.50 & 2.39 $\pm$ 3.23 & \\
 \hline 
\multirow{14}{2cm}{\centering MAGIC ($>80$ GeV)} &Pre-2017& \textbf{PL$_\mathrm{baseline}$} & $(2.04\pm0.63)\times 10^{-14}$&3.50&$2.23\pm0.17$&-& \multirow{2}{*}{\writeChitworow{9.6/12}}\\ 
 &Fall 2017& PL$_\mathrm{baseline}$ + \textbf{PL$_\mathrm{binary}$} & $(1.65\pm0.33)\times 10^{-12}$&0.70&$2.61\pm0.18$&-& \\ 
\cline{2-8}
  &Pre-2017& \textbf{PL$_\mathrm{baseline}$} & $(2.20\pm0.64)\times 10^{-14}$&3.50&$2.17\pm0.26$&-& \multirow{2}{*}{\writeChitworow{4.8/11}}\\ 
 &Fall 2017& PL$_\mathrm{baseline}$ + \textbf{PLEC$_\mathrm{binary}$} & $(3.77\pm1.68)\times 10^{-12}$&0.70&$1.74\pm0.37$&$1.40\pm0.97$& \\
\cline{2-8}
 &Pre-2017& \textbf{PL$_\mathrm{baseline}$} & $(2.30\pm0.67)\times 10^{-14}$&3.50&$2.15\pm0.19$&-&\multirow{3}{*}{\writeChithreerow{4.4/15}} \\ 
 &Low State& PL$_\mathrm{baseline}$ + \textbf{PL$_\mathrm{binary}$} & $(9.84\pm3.41)\times 10^{-13}$&0.70&$2.57\pm0.26$&-&\\
 &High State& PL$_\mathrm{baseline}$ + \textbf{PL$_\mathrm{binary}$} & $(3.69\pm0.64)\times 10^{-12}$&0.70&$2.17\pm0.23$&-& \\
\cline{2-8}
 &Pre-2017& \textbf{PL$_\mathrm{baseline}$} & $(2.63\pm0.60)\times 10^{-14}$&3.50&$2.06\pm0.17$&-&\multirow{3}{*}{\writeChithreerow{3.0/14}} \\ 
 &Low State& PL$_\mathrm{baseline}$ + \textbf{PLEC$_\mathrm{binary}$} & $(5.11\pm3.61)\times 10^{-12}$&0.70&$1.55\pm0.61$&$0.58\pm0.33$&\\
 &High State& PL$_\mathrm{baseline}$ + \textbf{PL$_\mathrm{binary}$} & $(1.65\pm0.14)\times 10^{-12}$&0.70&$2.20\pm0.40$&-& \\ 
\cline{2-8}
& Pre-2017& \textbf{PL$_\mathrm{baseline}$} & \multicolumn{5}{c}{\multirow{3}{*}{\textit{Insufficient data to discriminate between PL and PLEC in High State.}}} \\
& Low State& PL$_\mathrm{baseline}$ + \textbf{PLEC$_\mathrm{binary}$} &  \\
& High State& PL$_\mathrm{baseline}$ + \textbf{PLEC$_\mathrm{binary}$} &  \\
\enddata
\end{deluxetable*}

\section{Discussion} \label{sec:discussion}

\psrbe\ is the second TeV gamma-ray binary system to be detected in which the nature of the compact object is clearly established. Non-thermal emission from these systems likely results from the interaction of the pulsar wind with the wind and/or disk of the Be star \citep{1997ApJ...477..439T, 1999APh....10...31K, 2013AnARv..21...64D}. Particles are accelerated at the shock which forms between the pulsar and Be star winds. These subsequently produce synchrotron emission from radio to X-ray bands and inverse Compton emission at TeV energies. Numerous competing factors play a role in creating and modulating the observed emission. These include the efficiency of inverse Compton production and the degree of photon-photon absorption, which both depend upon the geometrical properties of the system with respect to the line of sight and the intensity, wavelength and spatial distribution of target photon fields \citep{2005ApJ...634L..81B}. Additional factors include: the position of the pulsar in relation to structures in the stellar wind \citep{2018MNRAS.474L..22P}; the bulk motion and cooling of the post-shocked material \citep{2006A&A...456..801D}; the structure of the magnetic field around the star \citep{2005MNRAS.356..711S}; and the degree of magnetization of the pulsar wind, and its evolution with radial distance from the pulsar \citep{2009ApJ...702..100T}. Isotropized pair cascades, triggered by misaligned VHE photons which would not otherwise be observed, can also contribute to the emission \citep{1997AnA...322..523B,2014JHEAp...3...18S}. Finally, interactions with the material and radiation of a circumstellar disk, the defining feature of the Be stellar class, may also modulate the X-ray and gamma-ray fluxes \citep{2008MNRAS.385.2279S}.

Modeling the time-dependent broadband emission is therefore complex, and challenging. \citet{2017ApJ...836..241T} have presented a model which explains the increasing X-ray flux prior to periastron as the result of the radial dependence of the pulsar wind magnetization, and the X-ray suppression at periastron  due to Doppler boosting effects caused by bulk motion of the post-shocked flow, naturally leading to an emission light curve which is asymmetric with respect to periastron. A recently revised version of their model predictions is given in \citet{2018ApJ...857..123L}, and also in \autoref{fig:LC}. The model prediction matches the early part of the XRT light curve reasonably well, when scaled by a factor of 0.5, but is unable to reproduce the rapid brightening around MJD 58080, when \psrbe\ was at superior conjunction. This feature may be explained, at least in part, by interaction of the pulsar with the circumstellar disk of the Be star, which could be confirmed by observations of radio pulsations during the periastron passage. Alternatively, as discussed in \citet{2018MNRAS.474L..22P}, it may be caused by geometrical effects associated with the orientation of the stellar disk with respect to the pulsar's orbit. 

\citet{2018JPhG...45a5201B} also calculated gamma-ray emission from the system, including a detailed treatment of the pair cascades triggered by the absorption of primary gamma rays, and the subsequent production of inverse Compton emission. They do not calculate a detailed light curve, but conclude that the binary emission may dominate the overall VHE flux, becoming comparable to, or exceeding, the steady flux from \tevj\ for a few weeks around periastron and superior conjunction. The predicted elevated flux close to periastron of $\sim1.6\times10^{-12}\U{erg}\UU{cm}{-2}\UU{s}{-1}$ at $1\U{TeV}$ is similar to the high-state emission levels reported in this work.  We also note that the VHE efficiency (L$_{>200\U{GeV}}$/$\dot{\mathrm{E}}=1.4\%$) for \psrbe\ is approximately the same as that of PSR\,B1259-63 / LS\,2883. In contrast, the GeV efficiency of \psrbe\ is significantly lower than that of PSR\,B1259-63 / LS\,2883, which can exceed 100\% \citep{2018ApJ...857..123L, 2018ApJ...863...27J}.

A distinctive feature observed in the VHE light curve is a sharp flux drop around seven days after periastron, lasting just a few days. As noted in \citet{2017ApJ...836..241T}, a similar dip has been seen in the lightcurve of PSR\,B1259-63 / LS\,2883, which \citet{2017ApJ...837..175S} attributed to photon-photon absorption.  This effect is predicted to be strongest when both the interaction angle between the photons is optimal and when the gamma-ray photons pass through the densest photon field, which occurs around superior conjunction, 5--15 days after periastron for \psrbe. 

Based on the detailed sampling of the VHE and X-ray light curves reported here, coupled with the measurement of an unexpected low-energy spectral cutoff in the VHE low state, it is clear that the existing models will require significant revision. Analysis of the pulsar timing evolution over periastron will provide important additional input, including more accurate measurements of the system geometry. It will also allow for more sensitive searches for GeV emission in the \textit{Fermi}-LAT data, with the dominant magnetospheric emission from the pulsar removed by a temporally-gated analysis.

Finally, it is interesting to reconsider the properties of the steady VHE source, \tevj, in the light of these results. As noted in \citet{2014ApJ...783...16A}, if we assume that \tevj\ is the pulsar wind nebula of \psr, then \psr\ is one of the oldest and weakest pulsars with a nebula seen in both X-ray and VHE gamma rays. 
In a recent population study, \citet{2017arXiv170208280H} derive empirical relations between VHE luminosity and pulsar spin-down energy, and also between PWN radius and characteristic age. For \psr, these relations predict a radius of over $20\U{pc}$ (compared to a measured extent of $4.7\times2.0\U{pc}$), and a TeV luminosity (1 to 10 TeV) of $2\times10^{33}\U{erg}\UU{s}{-1}$ (compared to the measured value of $8\times10^{32}\U{erg}\UU{s}{-1}$). However, the measured properties of VHE PWN display a large intrinsic scatter, and the physical size of the nebula can be strongly modified by the local interstellar environment.
We conclude that \psr\ remains a plausible candidate for the power source driving \tevj\ and note that it may be worthwhile to search for extended TeV nebulae around other known TeV binary systems -- although the formation of \tevj\ may only be possible due to the exceptionally long orbital period and large eccentricity of the binary system, which allows \psr\ to spend much of its orbit effectively as an isolated pulsar. 

X-ray and gamma-ray monitoring of \psrbe\ will continue. PSR\,B1259-63 / LS\,2883 produces bright gamma-ray flares in the days and months after periastron, and it ejects rapidly moving plasma clumps generated by the interaction of the pulsar with the stellar disk \citep{2015ApJ...806..192P}. Similar phenomena may occur in the case of \psrbe. The ongoing observing campaigns therefore provide a rare opportunity to completely sample the high-energy behavior of this system around periastron, which will not be repeated until approximately 2067.

\acknowledgments
VERITAS is supported by the U.S. Department of Energy, the U.S. National Science Foundation, the Smithsonian Institution, and by NSERC in Canada. We acknowledge the excellent work of the support staff at the Fred Lawrence Whipple Observatory and at collaborating institutions in the construction and operation of VERITAS.

We acknowledge \textit{Fermi} and \textit{Swift} GI program grants 80NSSC17K0648 and 80NSSC17K0314.

The MAGIC Collaboration thanks the funding agencies and institutions listed in:

\noindent \url{https://magic.mpp.mpg.de/ack_201805}

\vspace{5mm}
\facilities{\textit{Swift}-XRT, VERITAS, MAGIC}
\software{astropy \citep{2013A&A...558A..33A}, ROOT \citep{ROOT}, XSPEC \citep{Arnaud96}}

\end{document}